Tailoring Metallo-Dielectric Structures for Super Resolution and Super-Guiding Applications in the Visible and Near IR Ranges

D. de Ceglia[1,2], M.A. Vincenti[2,1], M.G. Cappeddu[1,3], M. Centini[4], N. Akozbek[1], A. D'Orazio[2], J.W. Haus[5], M.J. Bloemer[1], M. Scalora[1]

[1]*Charles M. Bowden Research Facility, AMSRD-AMR-WS-ST, RDECOM, Redstone Arsenal, AL 35898-50003*

[2]*Dipartimento di Elettrotecnica ed Elettronica, Politecnico di Bari, Via Orabona 4, 70125 Bari, Italy*

[3]*Dipartimeno dei Materiali, Università di Roma "La Sapienza", via Eudossiana 18, I-00184 Rome, Italy*

[4]*INFM at Dipartimento di Energetica, Università di Roma 'La Sapienza', Via A. Scarpa 16, 00161 Roma, Italy*

[5] *Electro-Optics Program, University of Dayton, Dayton, OH 45469-0245, USA*

## ABSTRACT

We discuss propagation effects in realistic, transparent, metallo-dielectric photonic band gap structures in the context of negative refraction and super-resolution in the visible and near infrared ranges. In the resonance tunneling regime, we find that for transverse-magnetic incident polarization, field localization effects contribute to a waveguiding phenomenon that makes it possible for the light to remain confined within a small fraction of a wavelength, without any transverse boundaries, due to the suppression of diffraction. This effect is related to negative refraction of the Poynting vector inside each metal layer, balanced by normal refraction inside the adjacent dielectric layer: The degree of field localization and material dispersion together determine the total momentum that resides within any given layer, and thus the direction of energy flow. We find that the transport of evanescent wave vectors is mediated by the excitation of quasi-stationary, low group velocity surface waves responsible for relatively large losses. As representative examples we consider transparent metallo-dielectric stacks such as Ag/TiO2 and Ag/GaP and show in detail how to obtain the optimum conditions for high transmittance of both propagating and evanescent modes for super-guiding and super resolution applications across the visible and near IR ranges. Finally, we study the influence of gain on super-resolution. We find that the introduction of gain can compensate the losses caused by the excitation of surface plasmons, improves the resolving characteristics of the lens, and leads to gain-tunable super-resolution.

## 1. INTRODUCTION

It is widely known that a thin layer of silver can act as a super-lens in the ultraviolet range [1, 2] for TM-polarized light. The explanation is twofold: (i) silver operating below its plasma frequency shows negative values of relative permittivity, so that incident TM-polarized waves produced by a light source located near a flat silver film undergo negative refraction and focusing, once inside the metal layer, and a second time behind the exit interface; (ii) if the silver film is placed quite close to the near field of a sub-wavelength-sized light source or aperture, the evanescent components of the near field can couple to surface plasmons excited at the air-metal interfaces, and they contribute to image formation along with propagating modes. As a consequence, unlike the case of conventional lenses, which capture propagating modes only, in the former the image is not diffraction-limited, and contains information about the sub-wavelength features of the illuminated object. In Pendry's original proposal [1], which consisted of a single 40-nm thick silver layer, two 30nm apertures placed on an absorbing, infinitely thin screen separated by a 90nm center-to-center distance, and located 20nm away from the input interface of the film, were resolved at a point 20nm outside the exit interface of the silver layer. The performance of this particular lens is frustrated by the naturally low degree of transparency attainable in transition metals such as silver, gold, copper, etc. Even for relatively thin layers (e few tens of nanometers thick, perhaps less, depending on the wavelength), these materials are natural mirrors below their respective plasma frequencies, which vary according to the material. On the other hand, the same materials display an intrinsic, natural window of transparency in the neighborhood of the plasma frequency. But even in this regime the skin depth, defined as the distance at which an electromagnetic wave is attenuated down to 1/e of its input value, is extremely small with respect to the incident wavelength (typically ~$\lambda/50$-$\lambda/100$ or so, at optical wavelengths).

These limitations have led researchers to formulate alternative designs of super-resolving systems, operating in the ultraviolet range, principally based on metallo-dielectric (MD) multilayers that feature both metal and dielectric layers having thicknesses that are much smaller than the incident wavelength [3-6]. Although these strongly anisotropic, non-resonant composite structures can slightly improve light transmission compared to a uniform metal layer in terms of angular-frequency and k-vector domains, significant drawbacks remain: (i) reflectance accounts for considerable distortions of the light source (i.e. accentuated feedback between the source and the lens); (ii) transmittance bandwidth is relatively narrow, thus limiting device tenability. In the case of the so-called hyperlens [3], for example, the transmittance of propagating modes at the operating wavelength of 365nm for the stack in question (8 periods of Ag (35nm)/Al2O3 (35nm)) is ~3% at normal incidence; (iii) range of operation is limited to ultraviolet wavelengths.

In ref. [7], a symmetric $Ag/Si_3N_4$ structure was proposed to achieve all-angle negative refraction and amplification of evanescent waves. Even in that case the stated objective was the improvement of device response in the evanescent part of the spectrum, i.e. $k_x>k_0$. But disadvantages remain, which translate into low transmittance of propagating waves, which does not exceed 15%; large reflectance, in excess of 40%; and sharply peaked wave-vector transmission functions that are qualitatively similar to the transmission functions in Pendry's original single silver layer.

An efficient way to improve the transparency of transition metals below and far from the plasma frequency had already been suggested in references [8, 9]. The structures proposed in those works consist of metallo-dielectric photonic band gap (MD-PBG) configurations, composed of alternating metallic and dielectric layers, where dielectric layers thickness can be on the order of the incident wavelength, so that resonance tunneling can occur. This kind of MD-PBG structure amounts to a system of strongly coupled Fabry-Perot cavities, where interference, resonant tunneling, and field localization effects combine to yield highly desirable properties: (i) significantly boost transmittance well above 50%; (ii) broaden the spectral width of the transparency window to cover the entire visible range; (iii) lower reflectance to near-zero values across the transmission window; and (iv) achieve tunability across the visible and near IR ranges and beyond. An important feature of these quasi-periodic structures is the optimization of the transmittance function provided by the addition of what amount to anti-reflection (AR) coatings. Typical AR coatings used to apodize and significantly augment the transparency of MD-PBG structures consist of dielectric layers half the thickness of the dielectric layers constituting the unit cell of the periodic structure. It has been demonstrated, theoretically and experimentally, that MD-PBGs with AR coatings are characterized by transmittance levels well above 50% across the entire visible range and beyond, using considerable amounts of metal with overall thicknesses reaching several hundreds of nanometers [8, 9].

The present effort represents the continuation of two previous contributions on the subject, where sub-wavelength focusing [10] and super-resolution [11] were separately reported in transparent, metallo-dielectric stacks in the visible range. In references [10] and [11] two distinct operational ranges were identified: (i) large negative refraction angles of the total electromagnetic momentum vector inside the lens; this regime is characterized by sub-wavelength focusing near the diffraction limit, and by the relative absence of surface plasmons and evanescent modes; and (ii) the manifestation of super-resolution, which is usually accompanied by the transport of evanescent waves and surface plasmon excitation. Here, we detail a few of the propagation effects that characterize the super-resolving regime. We analyze light localization effects and the role played by the evanescent components of the source field, showing how the information carried by these modes is transferred to the image plane through the transparent metal stack when a single and multiple slits are present.

Our results may be summarized as follows. The typical structure that we consider is designed in a way that limits diffraction effects: the local negative refraction of the Poynting vector (momentum density) inside the metal layers [10] is almost precisely balanced by the normal refraction process inside the dielectric layers, and waveguiding occurs. The *total* momentum vector thus acquires a large longitudinal component, and a small transverse part. This small transverse momentum component then couples to a quasi-stationary plasmon, having a transverse group velocity of order $10^{-2}$-$10^{-4}c$, depending on tuning conditions and the precise nature of the source. The excitation of the surface plasmon is then responsible for the transport of evanescent modes. A k-space analysis of the longitudinal Poynting vector reveals that evanescent modes significantly contribute to energy flow inside the stack, and that for apertures that are much smaller than the incident wavelength most of the energy resides with evanescent wave vectors. The exact amount of energy, its distribution and flow ultimately depend on the source, i.e. whether it is a single- or multiple-slit system, and the nature of the aperture.

In ref.[10] it was shown that the presence of metal within the multilayer stack and the achievement of large negative refraction angles do not automatically guarantee transport of evanescent modes and/or super-resolution, even if the stack is placed in close proximity to the aperture(s). In fact, the onset of the super-resolving state does not depend on the particular dielectric material juxtaposed to the metal layer having a particularly high index of refraction, or the local momentum density and direction. Rather, the key to the efficient transport of evanescent modes and super-resolution depends on the onset of the waveguiding mechanism, i.e. the establishment of a mostly longitudinal, total momentum vector having a transverse component small enough to couple with and excite surface waves. These particulars are elucidated in Fig.1, where we provide a pictorial view of the two distinct regimes that one may access using the same transparent metal stack: The negative refraction of the total momentum vector is associated with sub-wavelength focusing with resolution above the diffraction limit, while waveguiding is associated with super-resolution and transport of evanescent modes.

The calculations and results that we report here were performed and obtained in a two-dimensional simulation grid using two independent, mutually supportive numerical methods that make no approximations, and yield similar results: (i) a finite-difference, time-domain (FDTD) method, whose details are discussed in reference [12], for example, with the inclusion of dispersive behavior in both the electric and magnetic responses, and (ii) a time-domain, fast Fourier transform-based beam propagation method (FFT-BPM) described in refs. [10] and [13]. The paper is organized as follows. In the next section we show the propagation properties of MD-PBG structures by first analyzing the transmission, reflection, and absorption characteristics of the typical structure that we consider. Then we focus on the waveguiding abilities of the structure. The waveguiding effect that we describe occurs in a direction normal to the length of the layers. We then examine the full-width at half-maximum (FWHM) of the electric and magnetic fields, as well as the Poynting vector at the exit interface of the multilayer. These fields are studied as functions of the cross-section of the TM-polarized light source incident on the input interface of the structure. Then we will illustrate how the waveguiding performance is altered as a function of the incident wavelength. In particular, we show the peculiar transport of energy associated with the evanescent components of the input field. We then propose and discuss two realistic MD-PBG structures, where the metal is still silver and the dielectric materials are GaP and $TiO_2$, for applications in two different wavelength ranges. Finally, we show that while the super-resolving regime is characterized by large losses due to the excitation of surface plasmons, doping the dielectric layers with a gain material can offset these losses and lead to gain-tunable super-resolution.

**Extreme Sub-Wavelength Light Channels in Resonant Metallo-Dielectric Structures**

The first structure we consider is sketched in Fig.2, and is a generic stack composed of a Ag/X multilayer, whose elementary cell has a 26-nm-thick Ag layer, and a 36-nm-thick layer of material X, coated by two AR coating layers of material X. For the moment X is assumed to be a dispersion-free, loss-less dielectric material with refractive index n=4, a restriction that will be removed later with minimal impact on both qualitative and quantitative aspects of our results. The dispersion profile of silver is found in Palik's handbook [14]. The plane-wave transmittance of this system is calculated with the transfer matrix method. In Fig.2 the transmission spectrum is plotted along with the reflection and absorption profiles, for a TM-polarized wave incident at 45° with respect to the *z*-direction. The frequency-domain behavior shown for this kind of metallic structure is the main reason for adopting the term *transparent metal*. In fact, the stack exhibits high transmittance and near-zero reflectance in a wavelength range that covers a large part of the visible spectrum, from 480nm to 760nm, in spite of the inclusion of 132nm of silver. Moreover, changing the angle of incidence does not significantly modify the shape of these spectra thanks to the large bandwidths involved. One may thus conclude that the transmittance of TM-polarized propagating modes easily acquires values in the 40-60% range, for angles of incidence well beyond 45°.

To reproduce the light emitted by a small object having sub-wavelength features, we consider a rectangular-shaped slit in a Germanium-like substrate [10]. Since our aim is to provide theoretical guidance for as realistic experimental conditions as possible, in what follows we describe realistic apertures having typical depths between 100 and 200nm. Outside the slit(s) region, this material provides high-absorbance and reflectance across the visible range (for simplicity we assume the index of refraction of the substrate is n~4.2+2.2i in the entire range of operation), and simultaneously limits the excitation of modes propagating along its surface, so that interference effects due to reflections between the source and the structure adjacent to the substrate are minimized.

Given the dispersion relation $k_z^2 = k_0^2 - k_x^2$ for input light traveling along the *z* direction and diffracting/propagating on the *x*-*z* plane, the propagating components of a generic source are those modes with real values of the wave-number $k_z$. Then, in free space, $|k_x| \le k_0$. The modes with $|k_x| > k_0$ represent the evanescent components of the source spectrum, and if vacuum or other ordinary dielectric, non-metallic material is present on either side of the slit or source, these modes exponentially decay along the *z* direction. For example, the spot size produced by a 12nm rectangular slit, illuminated by a light source having incident wavelength λ=640nm, propagating down to a distance L=348nm (typical stack thickness) inside a high-index material (n=4) is roughly ~3μm. Evanescent components emerging from this aperture quickly and completely decay within a distance of approximately 20nm. This behavior confirms that we are operating in a hyper-diffractive regime (the Fresnel number $F = a^2/\lambda L \sim 10^{-5}$, where a=12nm). In order to guide the light emanating from such a small source, the stack should simultaneously be able to bend the propagating modes accordingly, avoid the rapid decay of the evanescent wave vectors, and transport them instead to the exit interface.

Let us now consider the multilayer stack depicted in Fig.2, and place it adjacent to a 12nm aperture. In Fig.3 we report the characteristic, time-averaged quantities derived using the FDTD method, obtained by averaging the electromagnetic field over several tens of optical cycles. This type of averaging corresponds to calculating the envelope function using the FFT-BPM. These two very different approaches yield nearly identical results. The shape of the Poynting vector $S_z$, and the electric and magnetic field intensities inside the structure show that this device confines and guides electromagnetic energy along the longitudinal direction, *normal* to the length of the material slabs. Given the extreme situations that we are considering, it is remarkable to see that there is significant light guidance through the entire length of the structure in the absence of side walls or other means of transverse confinement. Beginning with a slit 12nm wide, the FWHM of the longitudinal Poynting vector $S_z$ at the exit interface of the transparent metal is about 60nm. Thus diffraction effects are unambiguously inhibited, and the device acts as a super-waveguide for a slit ~λ/53 wide. This behavior is principally due to the fields' ability to access and become localized inside the metal layers, and thus to store energy and momentum there without the usual detriments of absorption [8-10]. Then, by combining the anomalous refraction process that goes on inside the metal with the normal refraction inside the dielectric layers, energy and momentum can be shaped and controlled in such a way that the average direction of energy flow inside the entire stack (calculated using the total momentum vector: $\mathbf{P}(t)=\int\limits_{z=-\infty}^{z=\infty}\int\limits_{x=-\infty}^{x=\infty}\frac{\mathbf{S}(x,z,t)}{c^2}\,dx\,dz=\int\limits_{z=-\infty}^{z=\infty}\int\limits_{x=-\infty}^{x=\infty}\frac{\mathbf{E}\times\mathbf{H}}{4\pi c}\,dx\,dz$ ) results at virtually zero degrees with respect to the axis of the rectangular aperture, for almost any angle of incidence. Inside the stack, contributions to the total electromagnetic momentum come from all possible k-vectors, including evanescent modes. Once the energy reaches the end of the stack, the wave recovers its initial propagation angle, and normal diffraction effects take place.

Even though the total momentum vector **P** refracts at near-zero angles inside the stack, displays a large longitudinal component, and is the determining factor in the successful excitation of surface waves and super-resolution, the local momentum density **S** also exhibits interesting dynamical characteristics. To be sure, following the temporal dynamics of short pulses traversing the stack offers remarkable glimpses into details of the interaction (such as how energy builds up and is absorbed inside the aperture itself and within the stack) that are not available in the steady state regime. In Fig.4 we show a vector-arrow diagram that represents the amplitude and direction of the energy flow $\mathbf{S}_z(x,z)$ in the continuous-wave excitation regime, for the 12nm aperture discussed above. The balancing act played by refraction between adjacent layers inside the stack is accompanied by energy circulation in proximity of the metallo-dielectric interfaces, where refraction alternates between anomalous (metal) and normal (dielectric). At these positions, the evanescent components couple to quasi-stationary, guided modes (surface plasmons), having low group velocity along the transverse direction. These evanescent modes strongly interfere with each other and with adjacent propagating modes, inducing rotation in the energy-flow that is generally more pronounced inside the stack when multiple slits are present. We note that the volume occupied by the aperture itself becomes a resonant cavity and filter for evanescent modes, and circulation is present there as well. Confirmation for this circulatory phenomenon, previously reported by Webb and Young [15] in similar structures, is given by the $S_z$ profile at the exit interface plotted in Fig.2, where the local momentum density can become negative.

Since one of our goals is to try to understand the dynamics of both propagating and evanescent modes, their distributions, and how they interact under realistic conditions, we now examine two representative examples. We first consider light emanating from a single 50nm slit (~λ/12), and impinging on the MD-PBG stack depicted in Fig.2. In Fig.5 we plot the longitudinal Poynting vector $S_{\mathbf{z}}(x,z)$ inside and just passed the stack. Although the guiding properties of the stack are obvious, the plot is also characterized by dark blue regions symmetrically located around the central guiding channel, where the momentum density takes on negative values, consistent with the Poynting vector depicted in Fig.3. In turn, $\mathbf{S_z}(x,z)$ may be analyzed further, and a Fourier decomposition immediately reveals its dependence on the transverse wave vector $k_x$; we plot the function $S_{\mathbf{z}}(k_x,z)$ in Fig.6. The figure shows that evanescent wave vectors become propagating inside the stack, and in fact carry most of the energy forward by coupling to the dissipative, quasi-stationary surface plasmons.

To visualize these processes, we now take a longitudinal cross section in the $S_{\mathbf{z}}(k_x,z)$ graph, and in Fig.7 plot the field amplitudes along with the corresponding Poynting vector for $k_x/k_0$~|3.2|. The oscillatory behavior and localization properties of the fields make them indistinguishable from propagating modes, at least inside the stack. For comparison, the inset of Fig.7 is a representation of the propagating mode $k_x/k_0$=0. Once the fields spill outside the stack they rapidly decay almost completely within a distance of approximately 10-20nm. The behavior of the specific evanescent mode chosen for illustration purposes is representative of the behavior of similar adjacent modes. However, the energy and momentum associated with a particular mode depend on the details of the source and ensuing interference effects. To illustrate this point we now consider two identical, 50nm slits (λ/12) having a 150nm center-to-center distance (λ/4). We place the stack of Fig.2 adjacent to the apertures, as shown in Fig.8, and recalculate the dynamics illustrated in Figs.(5-7). In Fig.9 we depict the longitudinal Poynting vector as a function of the transverse and longitudinal coordinates. Two resolved, guiding channels are clearly visible, with central depressions where the momentum density takes on negative values, and where strong rotation may occur. In the inset we show the cross sections of the Poynting vector, magnetic and electric field intensities at the exit of the stack, which display fringe visibilities of ~33%, zero, and 70%, respectively. These results clearly suggest that one should sample all the fields at the exit of the stack before assessing or assigning the degree of super-resolution, especially in view of the fact that experimental results hinge on the detection of the power flow.

A Fourier decomposition of the Poynting vector along the transverse coordinate is shown in Fig.10. This figure should be compared to Fig.6. In going from one to two apertures the longitudinal momentum density distributions and the associated flows have changed dramatically. While in the case of a single slit most of the flow associated with evanescent wave vectors was positive inside the stack, in the case of two slits both positive and negative components can easily be identified, a clear indication that evanescent modes are resonating inside the stack. Taking a cross section at $k_x/k_0 \sim |3.2|$ on Fig.10 reveals a field dynamics substantially similar to that depicted on Fig.7, except that now the fields combine with different phases to yield a mostly negative Poynting vector, an indication of energy circulation inside and just outside the stack. We show this in Fig.11, where we depict the dynamics of that particular mode beginning inside the substrate, where it originates, inside, and as it finally decays outside the stack. Again the mode can be seen to decay almost completely in free space, within a distance of 10-20nm.

There is remarkable dynamics of a fundamental nature that occurs inside the Ge substrate, where the mode also resonates and interferes with other modes before entering the stack. By choosing a realistic substrate, and by modeling the aperture as a realistic aperture, reveals additional dynamical features of light propagation phenomena at the sub-wavelength scale that are not present and remain unaccounted for in the usual treatments of sub-wavelength imaging [1, 4-6, 11, 15, 16]. The presence of a realistic scattering environment, where the aperture has both width and depth [10, 17-19], can significantly alter the dynamical range of the super-resolving regime, if not altogether inhibit it, as the density of evanescent modes can drastically change by specifying a somewhat arbitrary boundary condition on an unrealistic, infinitely thin absorbing layer. Since the issue of diffraction from realistic apertures is of fundamental importance in of itself [17-19], we choose to take it up its study separately.

**Aperture Variations**

We now analyze the characteristics of the fields as a function of aperture size, by calculating the FWHM of $S_z$, $|H_y|^2$ and $|E_x|^2$ at the exit interface for two different incident wavelengths, namely 690nm (Fig.12a) and 654nm (Fig.12b). The results provide

substantial confirmation that diffraction is efficiently inhibited by the multilayer stack over a wide range of wavelengths. The figures show two additional features of the structure: (i) the guiding ability of the stack depends on the wavelength of the source, especially for small apertures because they tend to re-distribute a significant portion of the available energy in the evanescent part of the k-vector spectrum and to foster absorption; (ii) adequate characterization of the guiding regime and actual resolving power should be done using all the components of the electromagnetic field, since at the exit of the stack the FWHM of $S_z$, $|H_y|^2$ and $|E_x|^2$ may be substantially different for each quantity (see for example Fig.9). Furthermore, this study provides a more precise idea of the minimum transverse size of the guiding channel permitted by the stack. For instance, using the incident wavelength of $\lambda$=690nm (Fig.12a), the FWHM for $S_z$ never dips below 57nm. At the same time, the minimum FWHM values for $|H_y|^2$ and $|E_x|^2$ are 67nm and 47nm, respectively. Oscillations around these values are observed at the wavelength $\lambda$=654nm, as depicted in Fig.12b. The sensitivity of the waveguide with respect to aperture variations is appreciable for apertures smaller than 50nm, and manifests itself in the oscillatory behavior seen in Fig.12a. While it becomes increasingly difficult to inhibit diffraction from point-like sources (<20nm or so), for larger apertures the FWHM is approximately linear, as the normal diffraction process takes hold without the excitation of evanescent modes.

Differences between small- and large-aperture regimes are substantial also in terms of energy transmitted and heat dissipated inside the stack. In the limit of a monochromatic wavelength of $\lambda$=600nm incident normally onto the stack, the transmittance is ~60%, while absorption losses are ~40%, as shown in Fig.2. When a sub-wavelength slit is drilled on the Ge substrate, which is then placed near the multilayer stack, the structure becomes sensitive to the presence of evanescent waves. In fact, the absorption versus transmittance ratio becomes gradually larger as slit size decreases, as more energy couples into the surface plasmon, and as an ever increasing portion of the incident energy is dissipated as heat by the stack. In Fig.13 we depict transmittance and absorption (left axis) through a stack having a single aperture, as a function of slit size, normalized with respect to the total energy that emerges from the aperture and is incident directly onto the stack. This normalization allows us to scale out the absorption of the aperture itself, which can easily be accounted for in our calculations. The absorption/transmission ratio is plotted on the right axis, and of course, this function is not sensitive to normalization. The curves thus suggest that the super-resolving regime (aperture size smaller than $\sim\lambda/2$) is characterized by surface plasmon excitation and heavy losses. We will return to the issue of losses in the last section, when we introduce a gain medium and study its effects on super-resolution. Suffice it to say here that for realistic, sub-wavelength apertures of order $\lambda/20$ evanescent waves and losses are clearly predominant, and the amount of energy absorbed by the stack via excitation of surface waves is approximately five times larger than the portion of transmitted power. These results largely confirm that the excitation of evanescent waves and the achievement of super-resolution come at a relatively high cost in terms of greatly reduced transmittance. For practical purposes, one should always weigh the benefits versus the expense one must incur in re-capturing the evanescent part of the field.

**Wavelength Variations**

Since the structure possesses a wide transparency band, the characteristics of the MD-PBG should be investigated through the entire visible spectrum. In Fig.14 the FWHM of the longitudinal Poynting vector $S_z$ is plotted as a function of the source wavelength for two fixed aperture widths, a=12nm and a=24nm. Under these highly diffractive conditions, the two curves reveal that the minimum FWHM occurs around λ=690nm, where the transmittance is approximately 45%. While positive-index lenses are able to mold the energy flow of propagating waves only, here the spot size is strongly influenced by the optical transfer function of the multilayer in the evanescent part of the spectrum. Predicting the wavelength at which the best resolution is achieved is not immediately obvious. Unlike non-resonant metallo-dielectric stacks proposed by other authors [3, 4, 6, 15], our lens cannot be regarded as an effective medium because of its field localization properties (e.g. Figs.7 and 11, and as discussed in ref.[10]), and it is not a perfect lens by any means. While in the effective medium limit both super-guiding and super-resolution come as a result of the large anisotropy induced by the alternation of positive and negative dielectric constants under non-resonant conditions, (the suppression of diffraction is obtained when the z-component of the effective dielectric constant tends to $\infty$ and the x-component is a small, positive number), the structures we propose are characterized mostly by field localization effects. Furthermore, the localization properties of the E and H fields are substantially different: the E field is more intense inside the dielectric layers, while the H field is more intense inside the metal layers. Under these conditions it is difficult to predict at what wavelength the spot size will be minimized.

**Surface Plasmons**

We have analyzed metallo-dielectric stacks as optical filters, and have studied their important peculiarities as functions of aperture size and incident wavelength. In this section we consider the optical transfer function $T(k_x,\lambda)$ of the stack discussed in the previous paragraphs, with the aim to show that its meaning is best exploited in terms of the guiding characteristics of the stack along the transverse direction, and the coupling mechanism of evanescent wave-vector components $k_x$ with the modes supported by the multilayer along this direction. In this context, the structure may be thought of as a multi-channel waveguide supporting multimode propagation. The optical transfer function thus represents a useful tool in the investigation of the dispersive properties of these modes, yielding information about phase and group velocities of transverse waves. A standard transfer matrix formalism has been used to derive the function $T(k_x,\lambda)$, whose shape is reported in Fig.15. The wide-band transparency of the lens in the *z* direction, both in the wavelength and the wave-vector domains, is directly observable by looking at the flat-shape behavior of the optical transfer function, with transmittance values around 60% in the propagation mode region $k_x<k_0$. The resonant MD-PBG thus provides strong coupling of evanescent modes with surface plasmons. In fact, at certain wavelengths one may have a high effective-refractive index (the ratio $k_x/k_0$ is precisely the effective refractive index along the transverse coordinate) and related low-group velocity. As a consequence, the effective propagating wavelength is small, giving the appearance of a state nearly frozen along the transverse direction, i.e. a quasi-stationary state. These excited states thus appear strongly localized inside the MD-PBG, as are the propagating modes, while they exponentially decay after traversing the exit interface. Thus interference between these modes and propagating modes gives rise to the vortex-like energy flow observable at the edges of the central region of the waveguide, where the energy is mostly channeled through the exit of the stack. Although the group velocity may be calculated directly from the dispersion relation suggested by Fig.15, another more direct and accurate way to proceed, which takes into account the nature of the source, is by integrating the equations of motion. For symmetric aperture systems, the net transverse energy flow is zero, as equal amounts of energy flow to either side of the aperture. The energy velocity along either the positive or negative transverse direction may be calculated simply as $V_e = \frac{\int_{x=0}^{\infty}\int_{z=0}^{L} S_x(z,x)\,dz\,dx}{\int_{x=0}^{\infty}\int_{z=0}^{L} U(z,x)\,dz\,dx}$, where $S_x$ and $U$ are the transverse Poynting vector and the energy density inside the stack, respectively, in the volume delimited by the limits of the integral. The magnitude of $V_e$ may be as small as $10^{-4}$c, depending on the carrier wavelength. The smallest transverse velocities coincide with the occurrence of smallest spot size.

**Realistic MD-PBG**

We now consider examples that will help shed light on some of the issues discussed in the previous paragraphs. The issue of the choice of the dielectric material is tightly linked to the structure capability of bending the incoming light. As addressed in ref. [10], the ratio between the real parts of the dielectric constants of the two materials and field localization characteristics play key roles in determining the operational regime of the metallo-dielectric stack. For example, in the structure proposed in that work large negative angles of refraction of the total momentum vector are achieved when the dielectric constant of the dielectric layers is sufficiently larger in magnitude than that of the metal. This means that focusing effects far from the exit interface of the stack can be obtained by using high-refractive-index materials. In contrast, small angles of refraction, super-guiding and super-resolution are achieved if the two dielectric constants are roughly comparable, but not necessarily identical, as in the perfect lens condition, as illustrated in Fig.1. As pointed out in the previous section, the large dispersion of metals in the visible range and field localization properties complicate any practical definition of a mathematical relation that would uniquely predict the best configuration of the structure in terms of geometry, i.e. the thicknesses of the metal and dielectric layers, and operational wavelength range. Nevertheless, in this section we give two examples of realistic, feasible stacks that operate in different parts of the visible range. The structures were designed to fulfill both the requirements of high transmittance in the visible range and good performance in terms of suppression of diffraction. The first stack is a 5.5 period Ag/GaP structure with layer thicknesses $d_{Ag}$=22nm and $d_{GaP}$=37nm, also reported in ref.[11]. AR coatings of GaP, which are 17nm thick, bound the multilayer. In Fig.16a we report the transmittance of this stack and the FWHM of the Poynting vector at the exit of the stack, corresponding to apertures 12nm and 24nm wide. Fig.16b, where the FWHM is plotted as a function of the aperture size at λ=582nm, reveals that this structure is still able to guide the light in sub-wavelength channels, in spite of the absorption losses and the smaller refractive index of the dielectric material ($n_{GaP}$@582nm~3.387). A cursory examination of Fig.16a shows that the same stack can either focus the incident light on a sub-wavelength scale (near the high frequency band edge, where spot size is large), or guide if it is tuned near 600nm.

The second example structure that we considered was designed to achieve super-resolution in the blue spectral region. It consists of 5.5 Ag/$TiO_2$ periods surrounded by AR coatings, with layer thicknesses of $d_{Ag}$=23nm and $d_{TiO2}$=50nm. In this case the refractive index of the dielectric material is about 2.4; hence the operational wavelength that maximizes the performance of the structure is expected to be smaller. Both these realistic structures are capable of imaging two sub-wavelength objects well below the Rayleigh limit. For example, using the $Ag/TiO_2$ stack, two 16nm apertures having a center-to-center distance of 120nm can be resolved with a visibility of ~60%. Alternatively, two 45nm aperture also separated by 120nm center-to-center distance can also be resolved with somewhat improved visibility.

**The Influence of Gain**

It has already been shown that in the context of non-resonant metallo-dielectric stacks the addition of an active medium has the potential of improving super-resolution [20]. The idea there was to circumvent material losses by introducing a material having a dielectric constant with exactly the opposite sign as the metal being used, in order to once again reclaim the perfect lens condition. In our case, we have seen that the operational range of super-resolution for any given stack is not tied to the perfect lens condition, and that losses are naturally suppressed by the resonant nature of the stack. Therefore, under realistic conditions metal absorption is generally not an encumbrance, and super-resolution is already obtained across a broad spectral range with relatively high transmittance. Instead, in resonant structures the issue is to see if the introduction of gain within the dielectric layers alters the relative localization properties of the fields inside the stack, and acts to either disrupt or improve the balance achieved between negative and normal refraction across the stack.

As in ref.[20], our results suggests that the presence of an active medium can improve the super-resolving properties of the lens: gain inside the dielectric layers boosts the electric fields there, with the corresponding magnetic fields growing inside the metal by a proportional amount. In Fig.17 we show the cross section of the Poynting vector at the exit of the stack, for the case of the double-slit/stack system depicted in Fig.(8-11), in the absence of gain, and then for two different gain coefficients. At 600nm, the dielectric constant of Ag is ε=-13.97 +i 0.93 [14]. The gain-less resolution of the two 50nm apertures having a center-to-center distance of 150nm yields a visibility for the longitudinal Poynting vector of roughly 33%. The figure suggests that by varying the imaginary part of the dielectric layer from 0 to 0.8 leads to a Poynting vector fringe visibility of approximately 46%.

The improvement that we report appears to be somewhat smaller than that implied in ref.[20], for example, in the context of an effective medium, even though we have used similar gain coefficients and dielectric constants. There may be multiple reasons for this behavior. For instance, we have seen that a more appropriate estimation of the resolving power of the lens should be made with respect to all the fields and power flow. Another significant difference may be due to the fact that we are considering realistic apertures, rather than the more idealized apertures placed on infinitely thin, absorbing screens. As we mentioned above, the imposition of such a boundary condition appears to be a standard theoretical approach in the context of sub-wavelength imaging, even though it is well-known that such a boundary condition does not satisfy all of Maxwell's equations [17-19]. In simple terms, a realistic aperture acts as a filter for propagating and evanescent modes alike; an aperture on an infinitely thin substrate tends to over-estimate the density of evanescent modes present at the aperture's exit, in some cases increasing the diffraction length by as much as a factor of two or more, depending on slit size. It is clear that this problem deserves special attention, and will be taken up separately.

**Conclusions**

We have described the extraordinary guiding properties of realistic, transparent, resonant MD-PBG structures, outlining their capabilities in terms of super-guiding and super-resolution across the entire visible range under hyper-diffractive conditions. The phenomena behind the strong lateral confinement provided by these structures are strictly related to the localization of the electromagnetic fields and the energy transport of both propagating and evanescent modes excited by sub-wavelength, realistic sources. The inhibition of diffraction is achieved over a wide spectral range thanks to the field localization properties peculiar to transparent metal stacks. Vortex-like patterns of the energy flux develop due to the presence of evanescent modes and their interference with propagating modes. We have proposed two practical designs based on Ag/GaP and Ag/$TiO_2$ multilayer stacks, demonstrating that super-resolution is preserved even in the case of lower-refractive-index dielectric materials. In the Ag/$TiO_2$ system, resolutions of the order of 0.25 λ are expected using sources only 0.036λ wide. Finally, we have seen that the introduction of gain helps the super-resolution process, and in effect renders the process gain-tunable within a realistic environment. There are additional issues that are worthy of exploration in the context of sub-wavelength imaging, namely diffraction from realistic apertures, that have finite width and depth. The characterization of this regime is multi-faceted, and clearly important for both basic and practical purposes. While we have discussed mainly negative refraction and waveguiding effects in the context of a super-resolving lens, our findings may also be relevant to the study of metals and near field optics in general, for applications in sub-wavelength light confinement, optical circuitry, sensing and plasmonic devices.

**Acknowledgments**: We thank Holger Fischer for suggesting the introduction of gain. We also thank the US-Army European Research Office for partial financial support.

## FIGURE CAPTIONS

**Fig.1:** Illustration of the different operating regimes that can be invoked in the same transparent metal photonic band gap structure. On the left we illustrate large negative refraction angles of the total momentum vector, and sub-wavelength focusing outside the stack takes place. On the right the total momentum refracts with a large longitudinal component, and super-guiding is accompanied by super-resolution.

**Fig.2:** Transmittance, reflectance and absorbance spectra of the 5.5 period Ag(26nm)/X(36nm) structure, coated with 18nm-thick layers of material X (n=4). The inset shows the sketch of the metallo-dielectric multilayer.

**Fig.3:** a) Spatial distributions of the electromagnetic field intensities $|H_y|^2$, $|E_z|^2$ and $|E_x|^2$, and the longitudinal Poynting vector $S_z$. The fields are averaged over 4 optical cycles. (b) Longitudinal Poynting vector $S_z$ profile at the exit interface $x$=378nm.

**Fig.4:** (a) Poynting vector **S** distribution on the *x*-*z* plane. The normalized arrows represent the energy flow inside the stack when a 24nm aperture carved on a Ge substrate is considered. The contour levels represent the magnitude of **S**. (b) **S** distribution in the neighbor of the exit interface.

**Fig.5:** Three dimensional view of the longitudinal Poynting vector from single 50nm aperture on a Ge substrate, inside and just outside the stack.

**Fig.6:** Fourier decomposition of the Poynting vector, $\mathbf{S}_z(k_x,z)$, depicted on Fig.5. This figure shows that channeling the light through a small aperture causes significant spectral broadening of the wave packet, with a good deal of energy concentrated in evanescent components.

**Fig.7:** Electric, magnetic and Poynting vector amplitudes inside the stack for the mode

$k_x/k_0=|3.2|$. The fields oscillate, resemble and are indistinguishable from propagating modes until the exit of the stack is reached, where all the fields are seen to decay rapidly.

**Fig.8:** Pictorial representation of the double-slit/stack system. For simplicity the stack is placed adjacent to the apertures.

**Fig.9**: Same as Fig.6, but for two apertures. Two channels are clearly visible, along with a central channel inside the stack where the Poynting vector acquires negative values. In the inset we offer a view of the cross section of the fields at the exit of the stack. Although the visibility of the longitudinal Poynting vector is ~33%, it is important to note that fringe visibility is practically zero for the magnetic field, and nearly 70% for the transverse E field. These results suggest that a proper assessment of the resolving capabilities of the lens should be made by sampling all the fields and power flow at the exit of the stack.

**Fig.10:** Fourier decomposition of the longitudinal Poynting vector of Fig.9. This figure should be compared with Fig.6, where we plot the same quantity but for a single aperture. In going from one to two slits the energy density and power flow have changed dramatically.

**Fig.11:** Same as Fig.7, but for two apertures. Here we also include the field profiles inside the aperture itself, where the mode is seen to resonate in the presence of effective impedance mismatch at both ends of the aperture. A comparison with Fig.7 also reveals that while the relative field amplitudes and localization properties remain almost unchanged, interference effects between adjacent evanescent modes, and between evanescent modes and propagating modes re-adjusts the phases of the fields to yield a completely different Poynting vector. Therefore, it is apparent that the evaluation of the dynamics of isolated evanescent modes, without the context of a realistic broadband source, is meaningless.

**Fig.12:** (a) FWHM of the electric field intensity, magnetic field intensity and longitudinal Poynting vector at the exit interface of the structure as functions of slit size. The source wavelength is 690nm; (b) Same as (a) at the source wavelength 654nm.

**Fig.13:** Transmission and absorption (left axis), and ratio of transmission/absorption (right axis) for the multilayer stack of Fig.2 placed near a single aperture of variable size placed on a Ge substrate. The transmission and absorption curves depicted on the left axis are normalized with respect to the energy that emanates from the exit of the aperture. It is clear that small apertures favor the onset of large absorption, as evanescent mode and surface plasmons are excited.

**Fig.14:** FWHM of the longitudinal Poynting vector plotted as a function of wavelength for two aperture sizes (12nm and 24nm). The Fresnel number for the two apertures differs by a factor of 4. Nevertheless, there is a region near 690nm where spot size is nearly constant, as a result of favorable super-guiding conditions.

**Fig.15:** Dispersion relation for the stack depicted in Fig.2.

**Fig.16:** (a) FWHM of the longitudinal Poynting vector at the exit interface of the Ag/GaP structures described in the text, plotted as a function of the wavelength for two aperture size values (12 and 24nm). The blue curve depicts the transmittance of the structure at normal incidence. (b) FWHM of the electric field intensity, magnetic field intensity and the longitudinal Poynting vector versus aperture size at the source wavelength 582nm. (c) Same as (a) but for the Ag/$TiO_2$ stack; (d) Same as (c) for an incident wavelength of 440nm.

**Fig.17:** Transverse profiles of the longitudinal Poynting vector at the exit of the stack with and without gain. The introduction of an active material with parameters not necessarily close to the perfect lens condition can improve the super-resolving properties of the stack. The visibility of the Poynting vector improves from 33% (no gain) to ~ 46% when Im(ε)= -0.8.

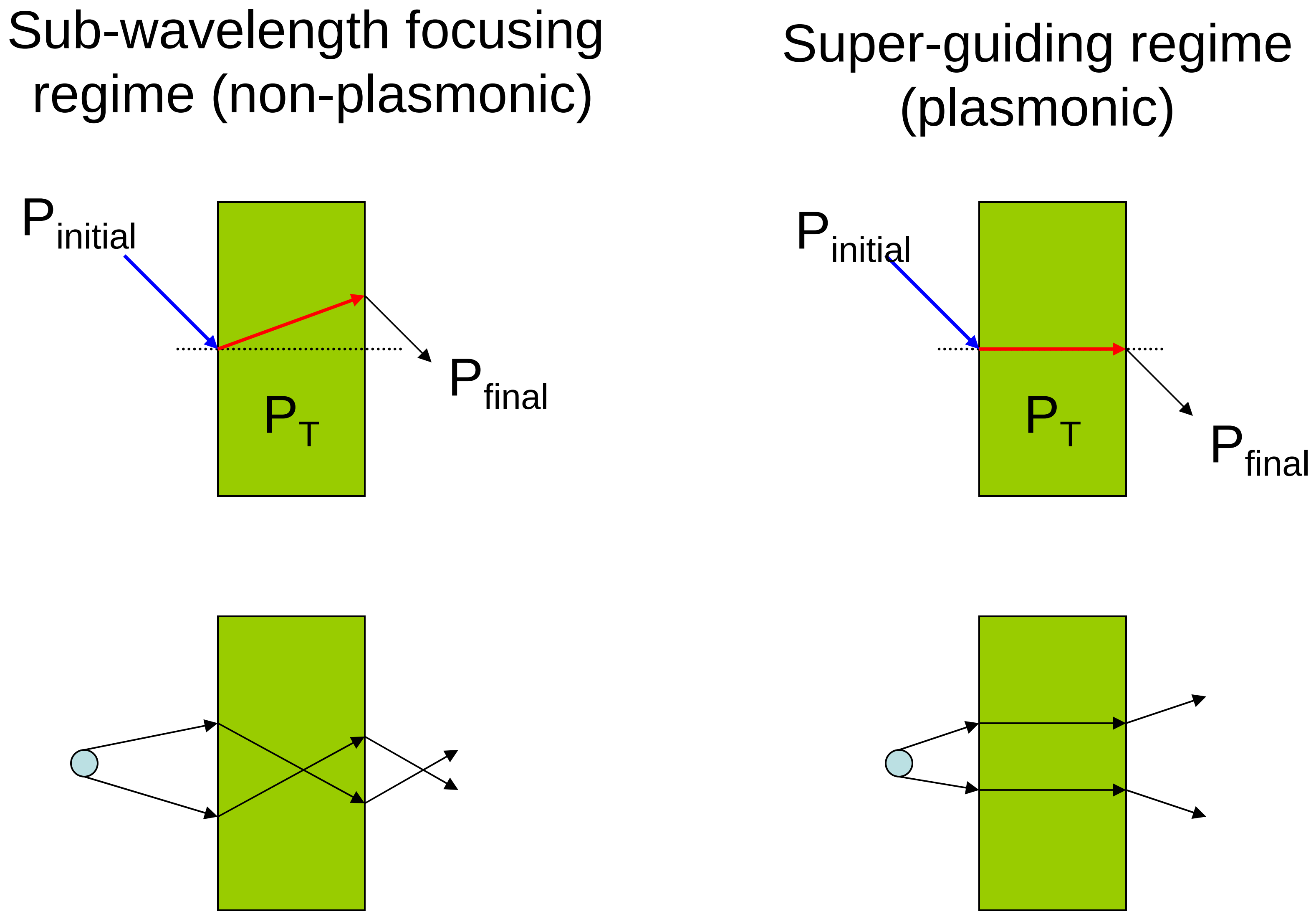

**Fig.1**

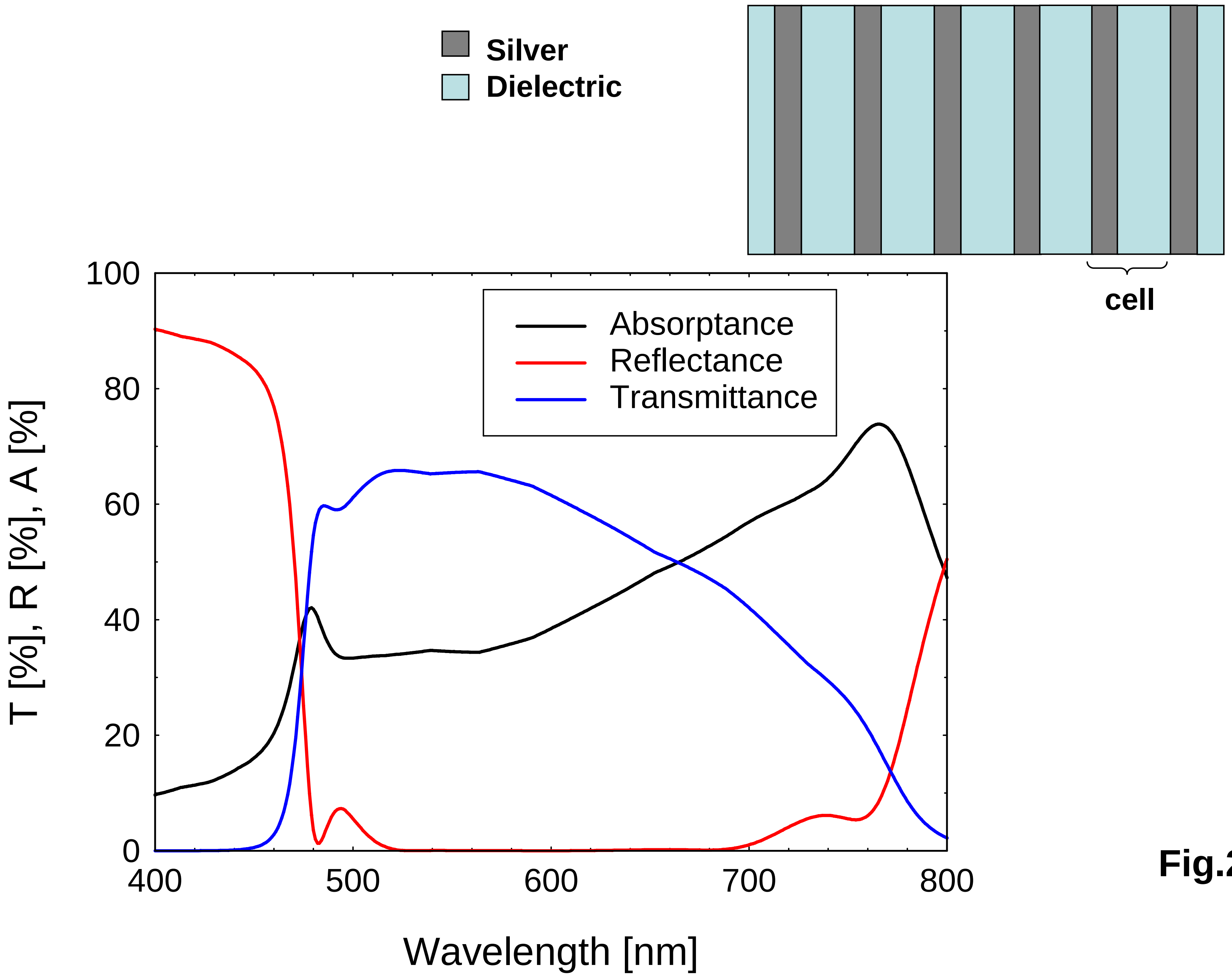

Fig.2

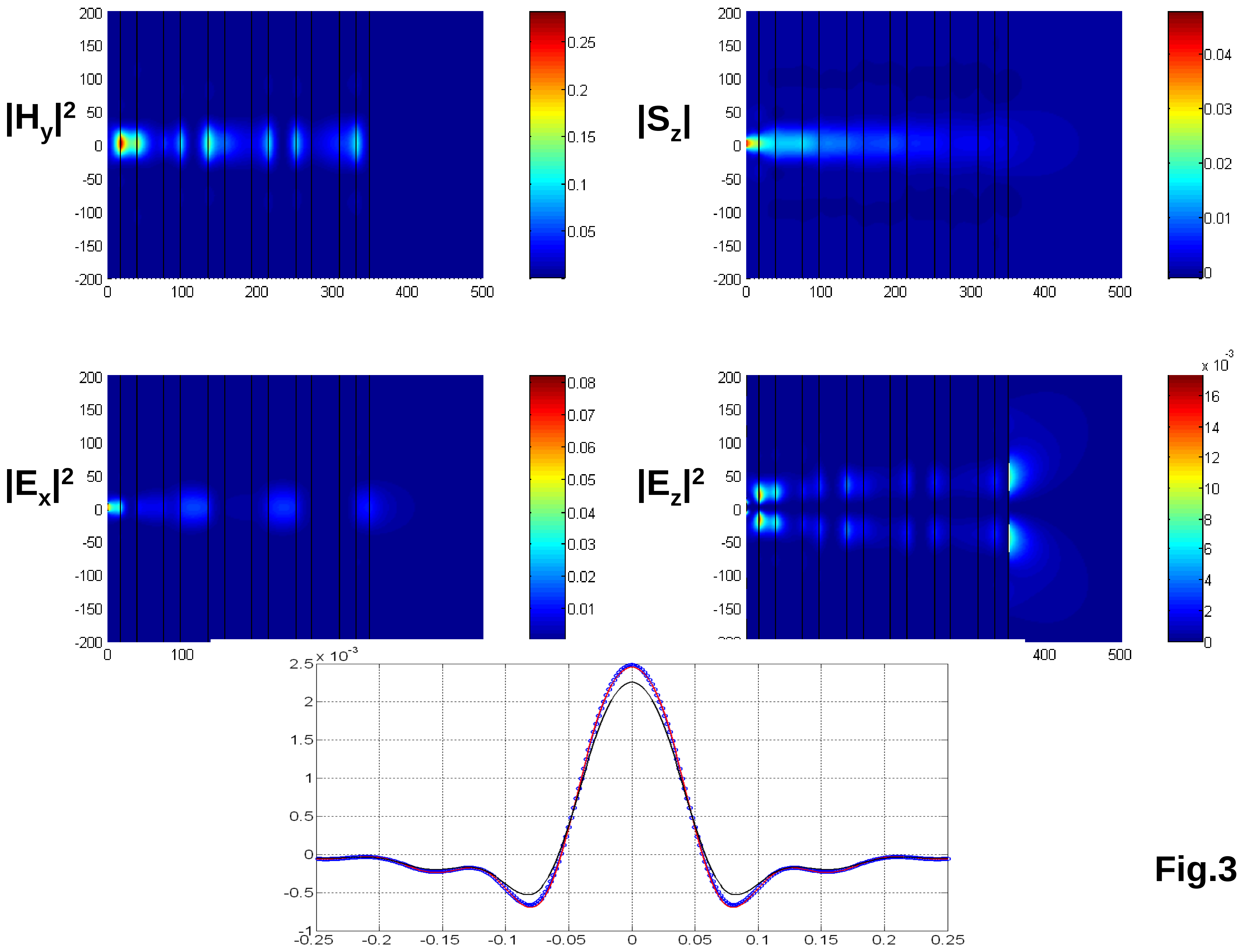

**Fig.3**

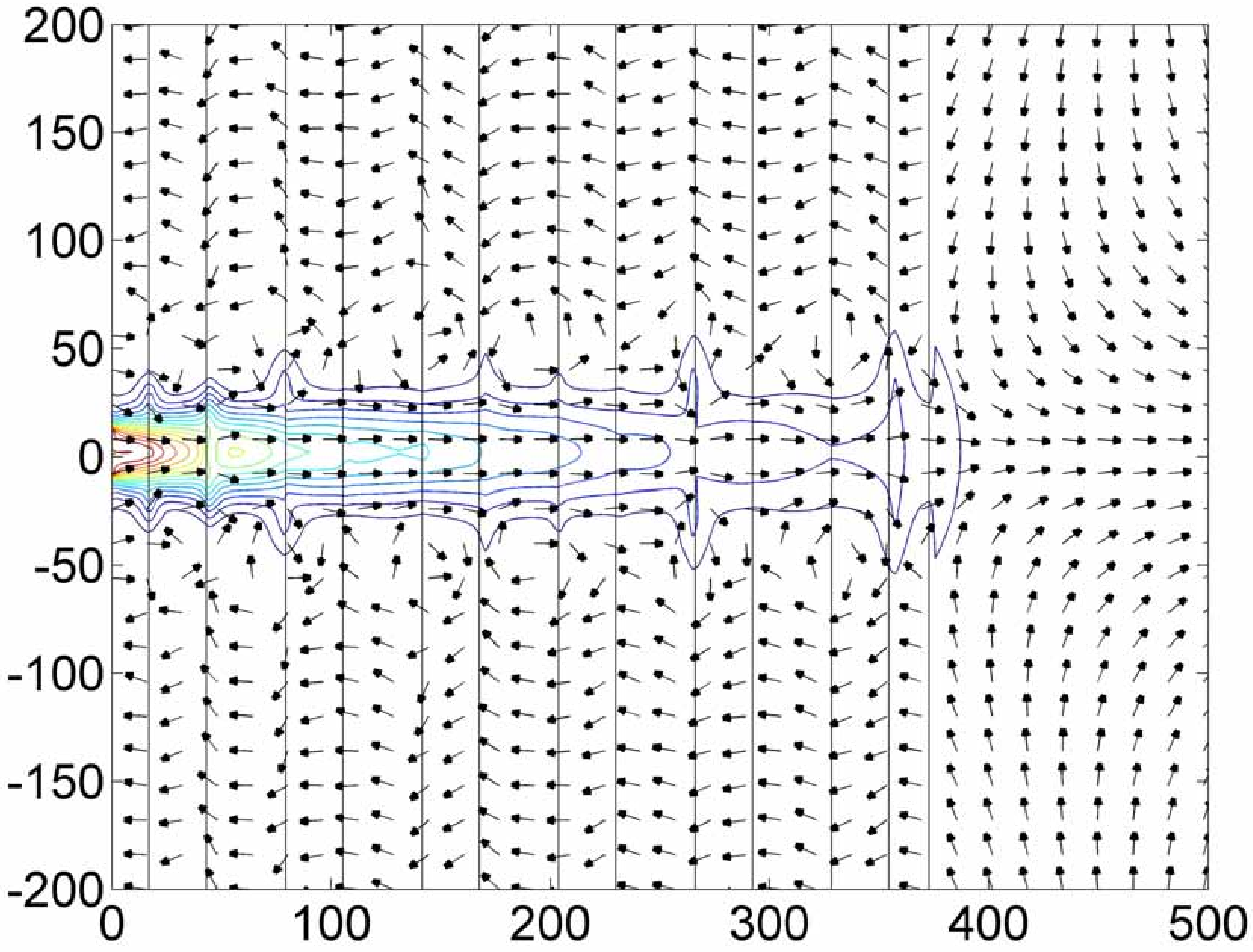

Fig.4

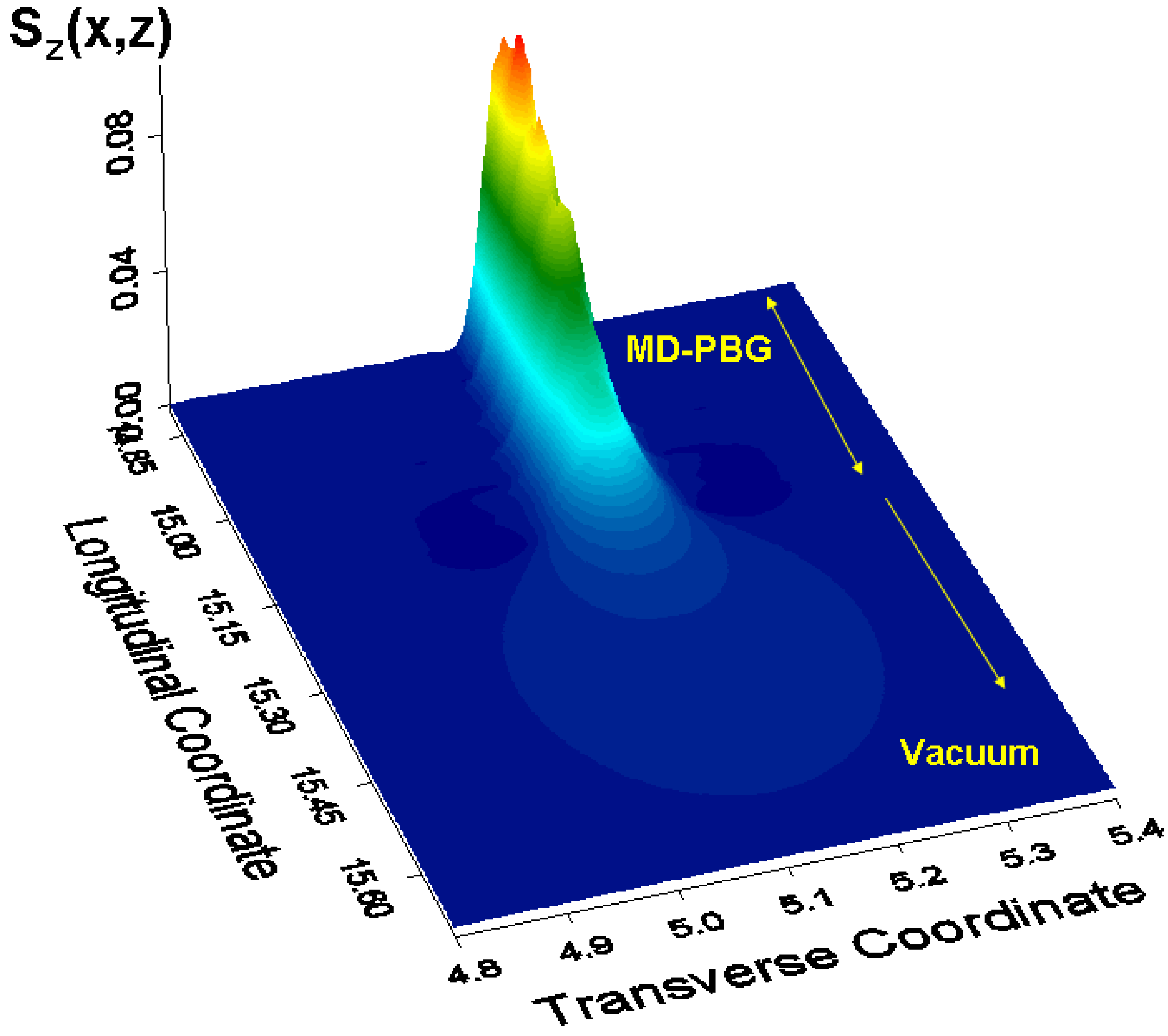

Fig.5

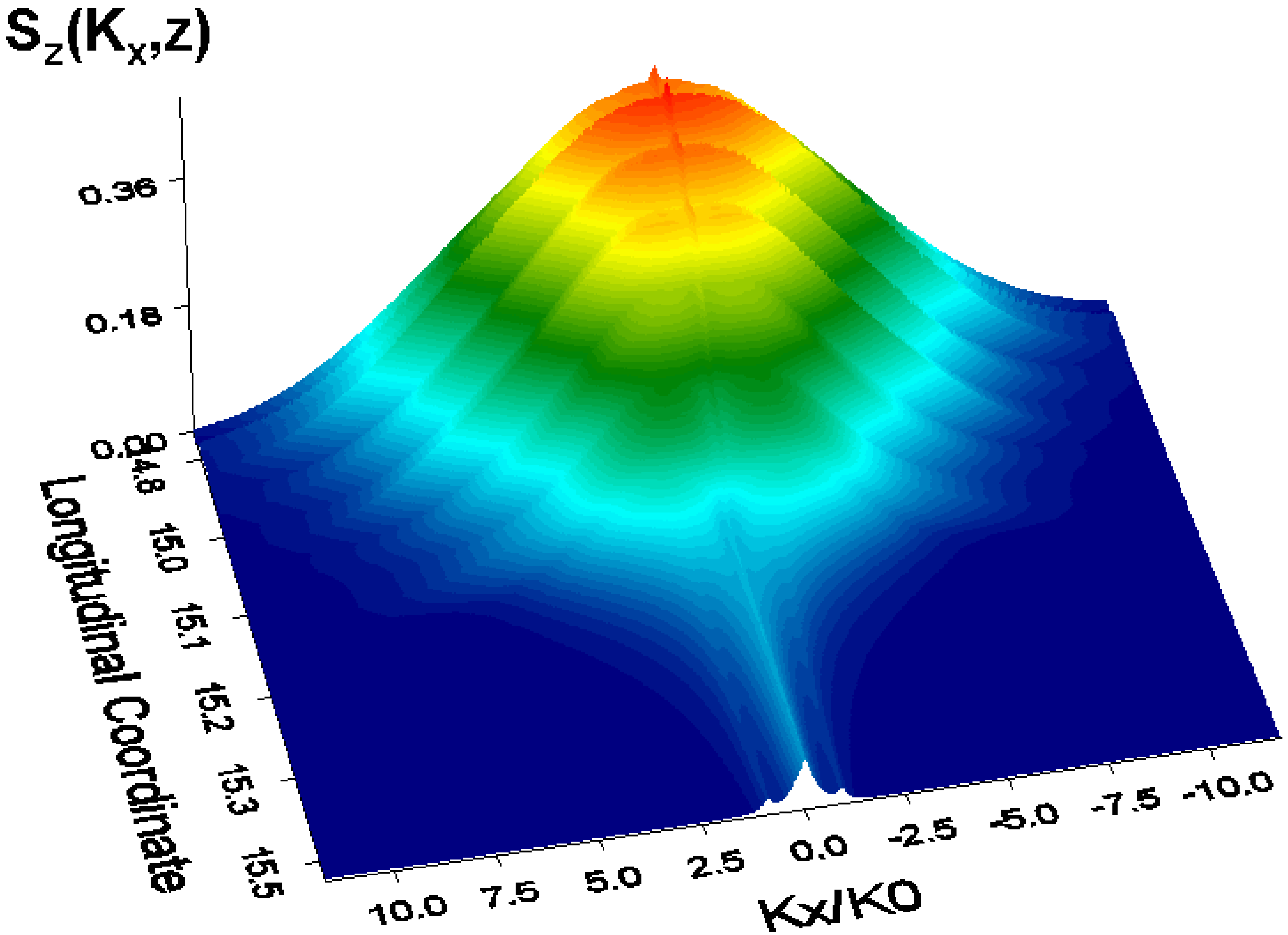

Fig.6

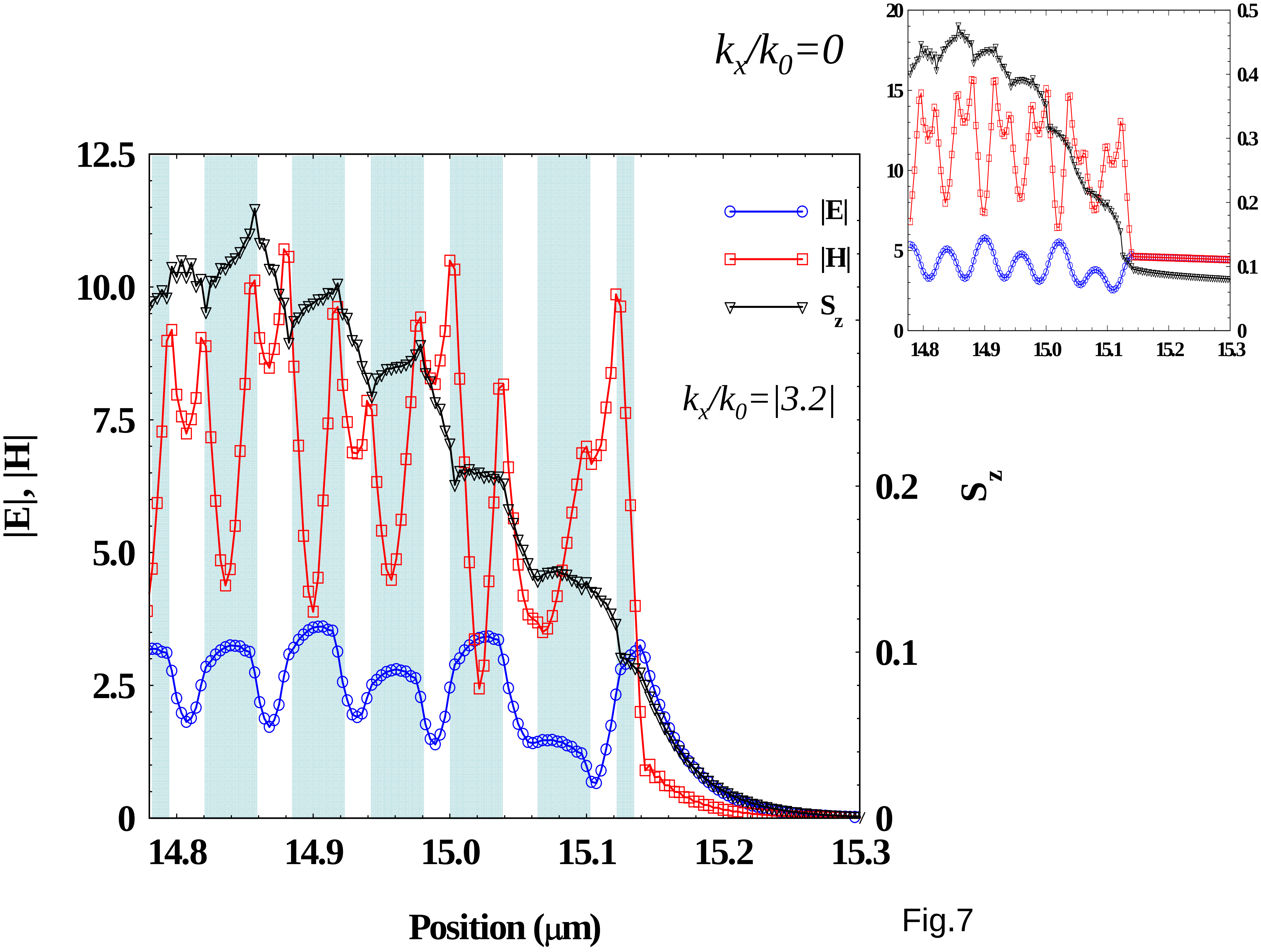

$k_x/k_0=0$
$k_x/k_0=|3.2|$
|E|
|H|
$S_z$
12.5
10.0
7.5
5.0
2.5
0
0.2
0.1
0
14.8
14.9
15.0
15.1
15.2
15.3
|E|, |H|
$S_z$
Position (μm)
20
15
10
5
0
0.5
0.4
0.3
0.2
0.1
0

Fig.7

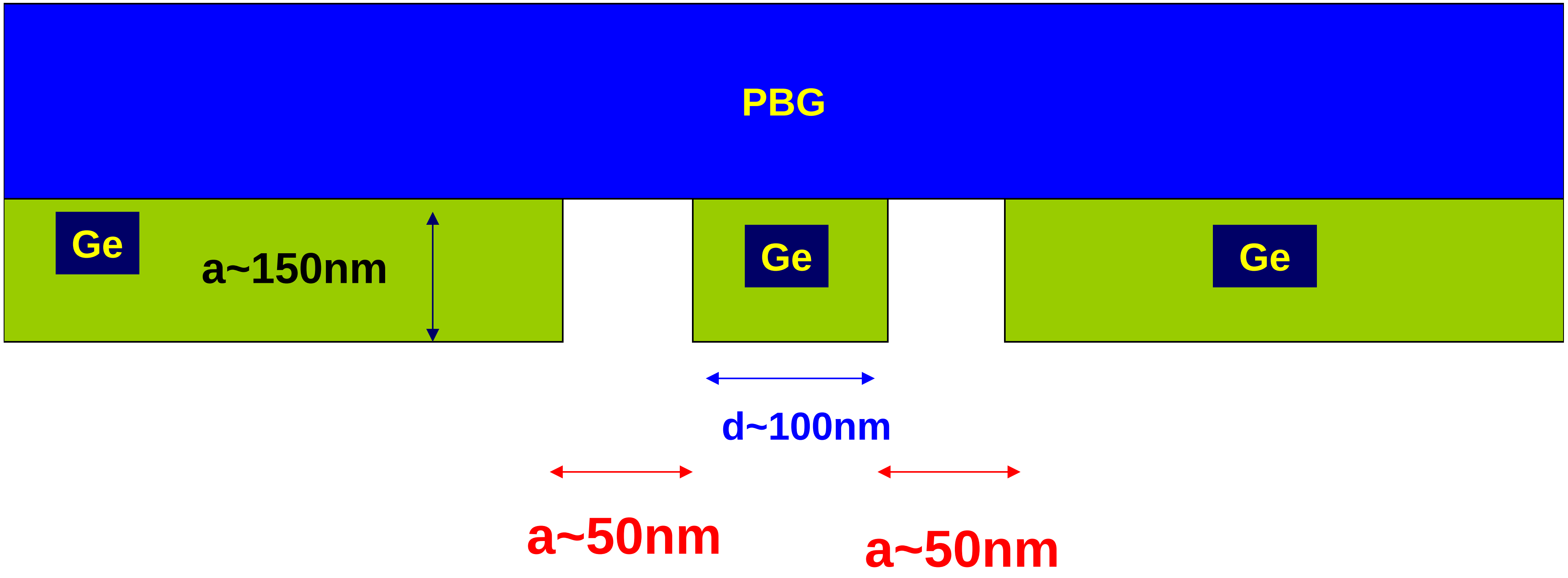

Fig.8

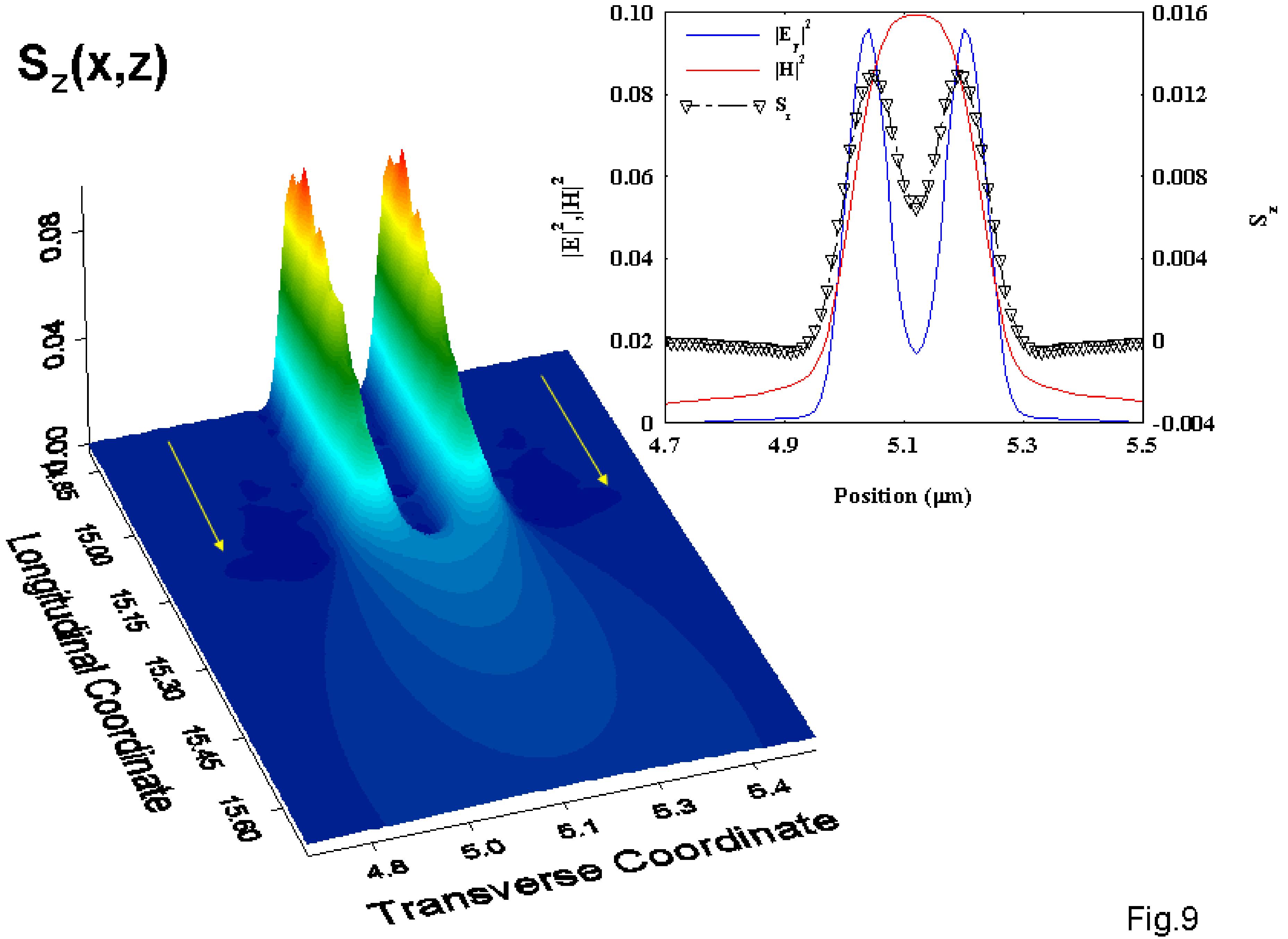

Fig.9

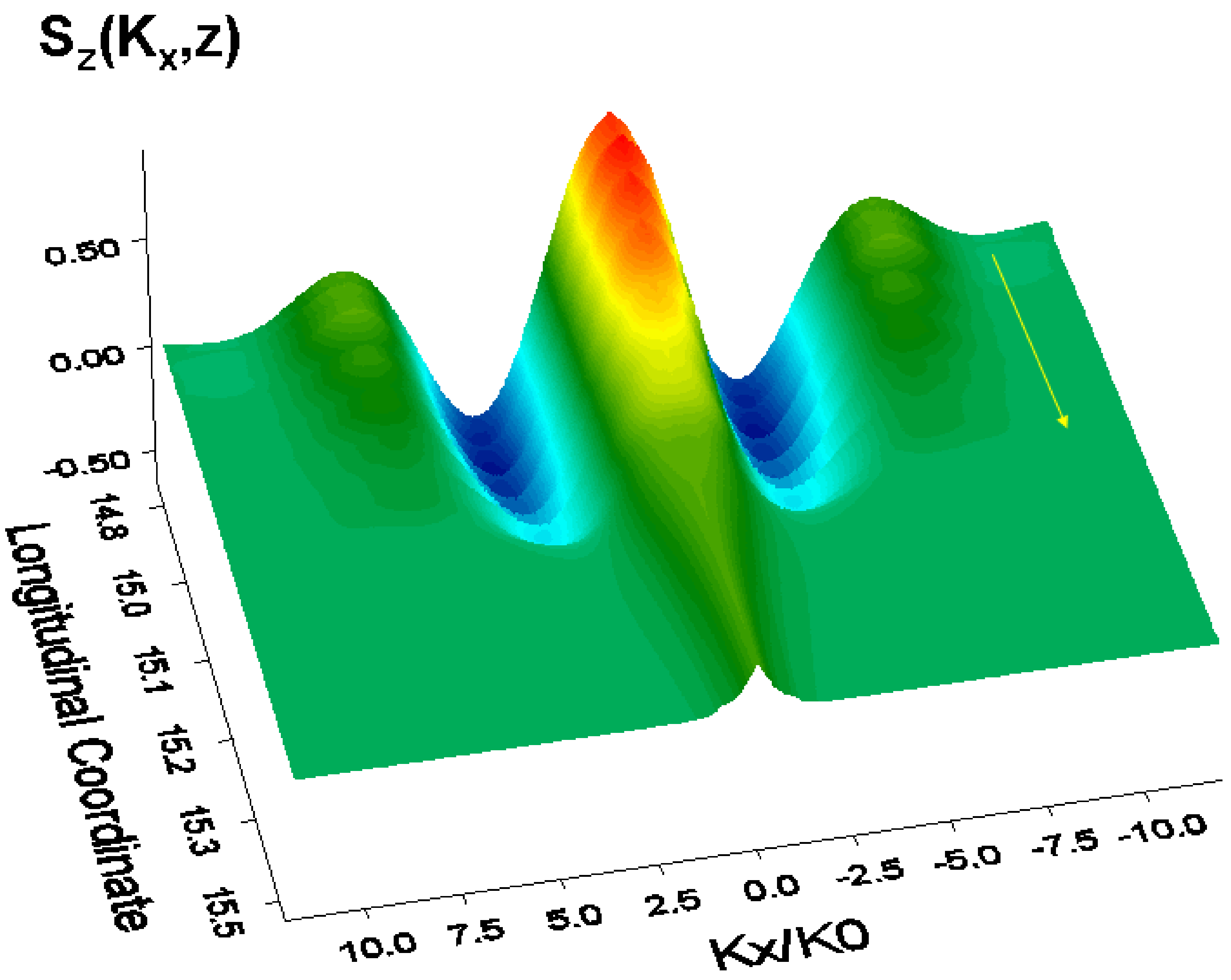

Fig.10

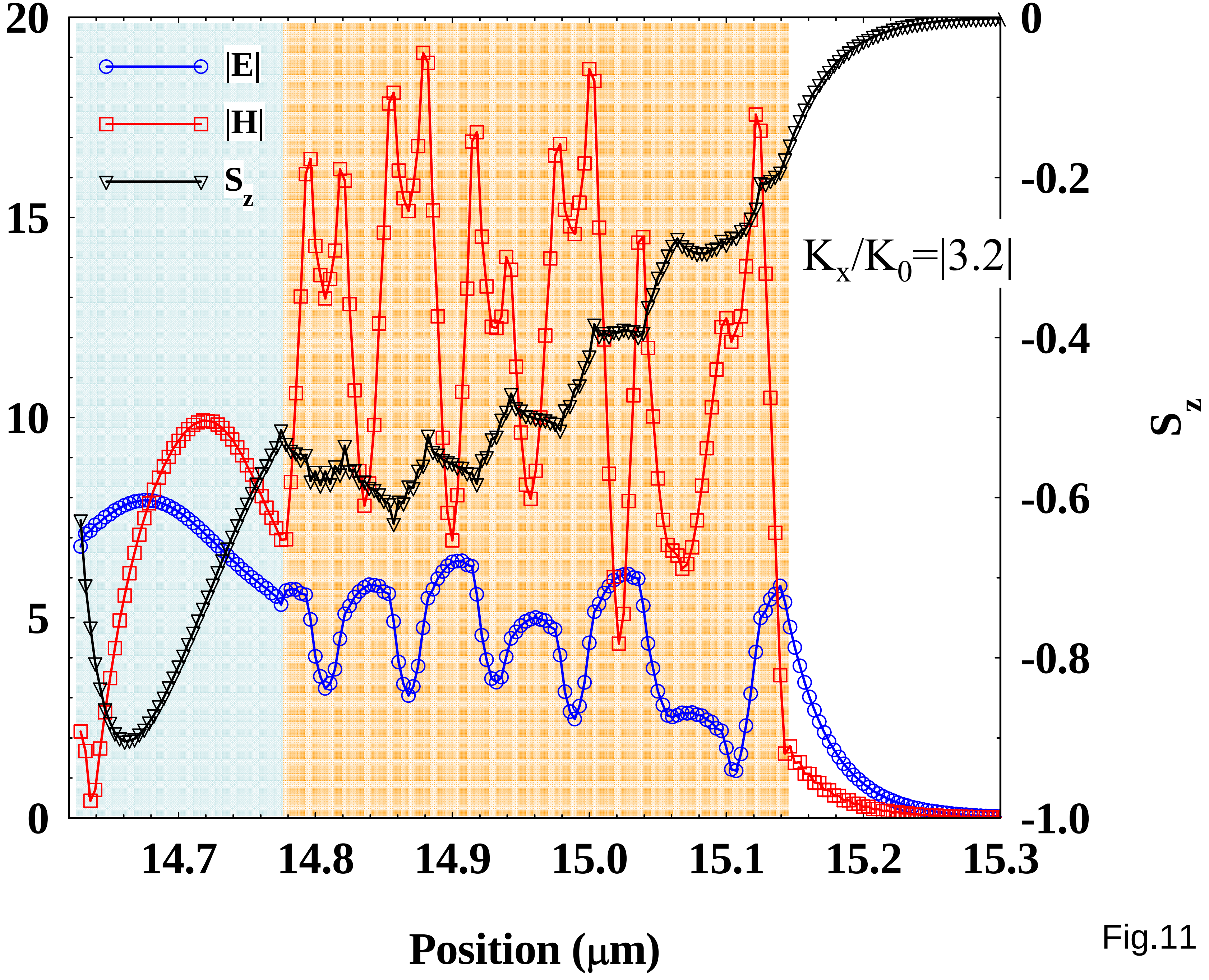

20
15
10
5
0
|E|
|H|
S_z
K_x/K_0=|3.2|
14.7
14.8
14.9
15.0
15.1
15.2
15.3
0
-0.2
-0.4
-0.6
-0.8
-1.0
S_z
Position (μm)

Fig.11

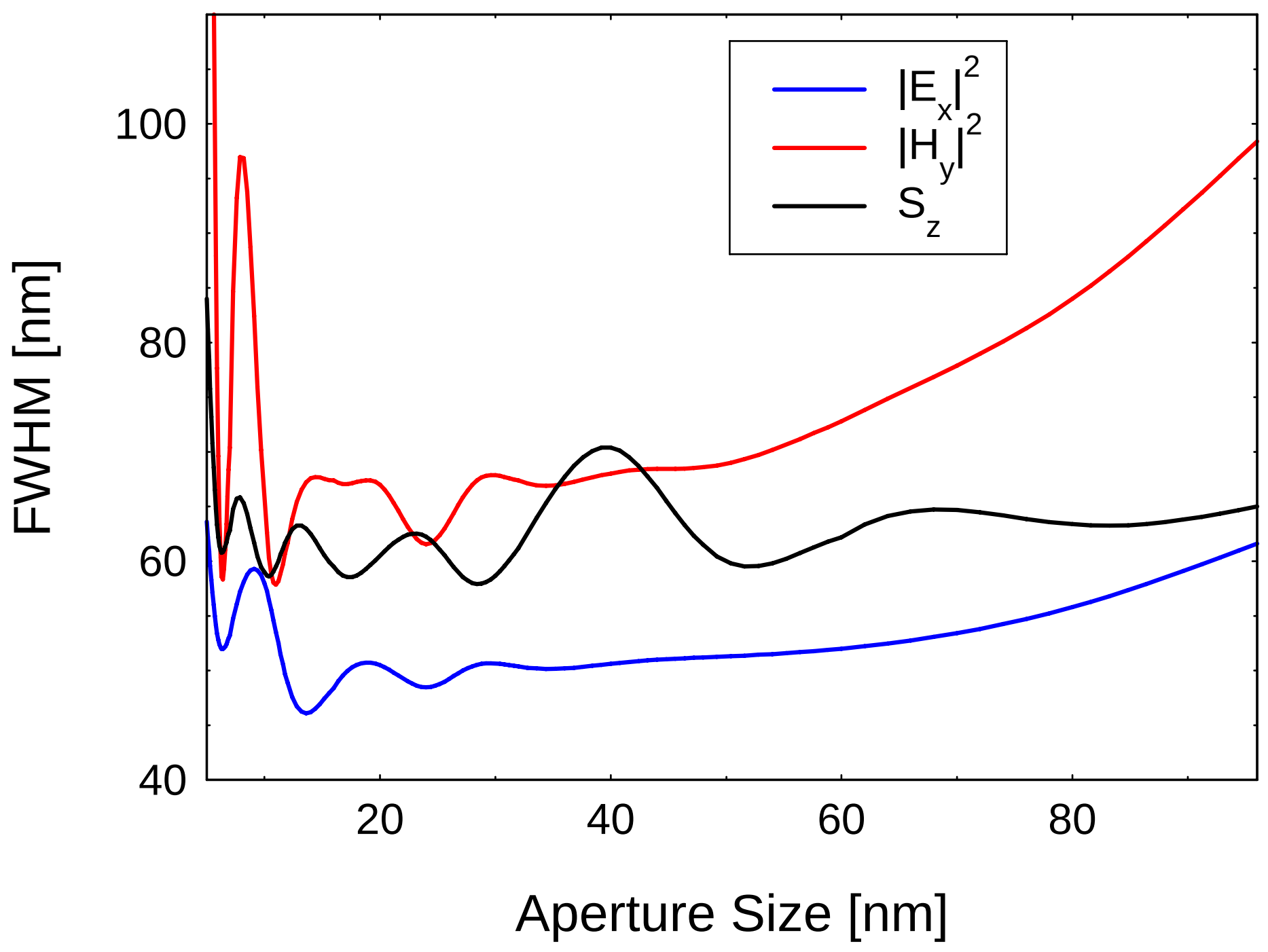

Fig.12a

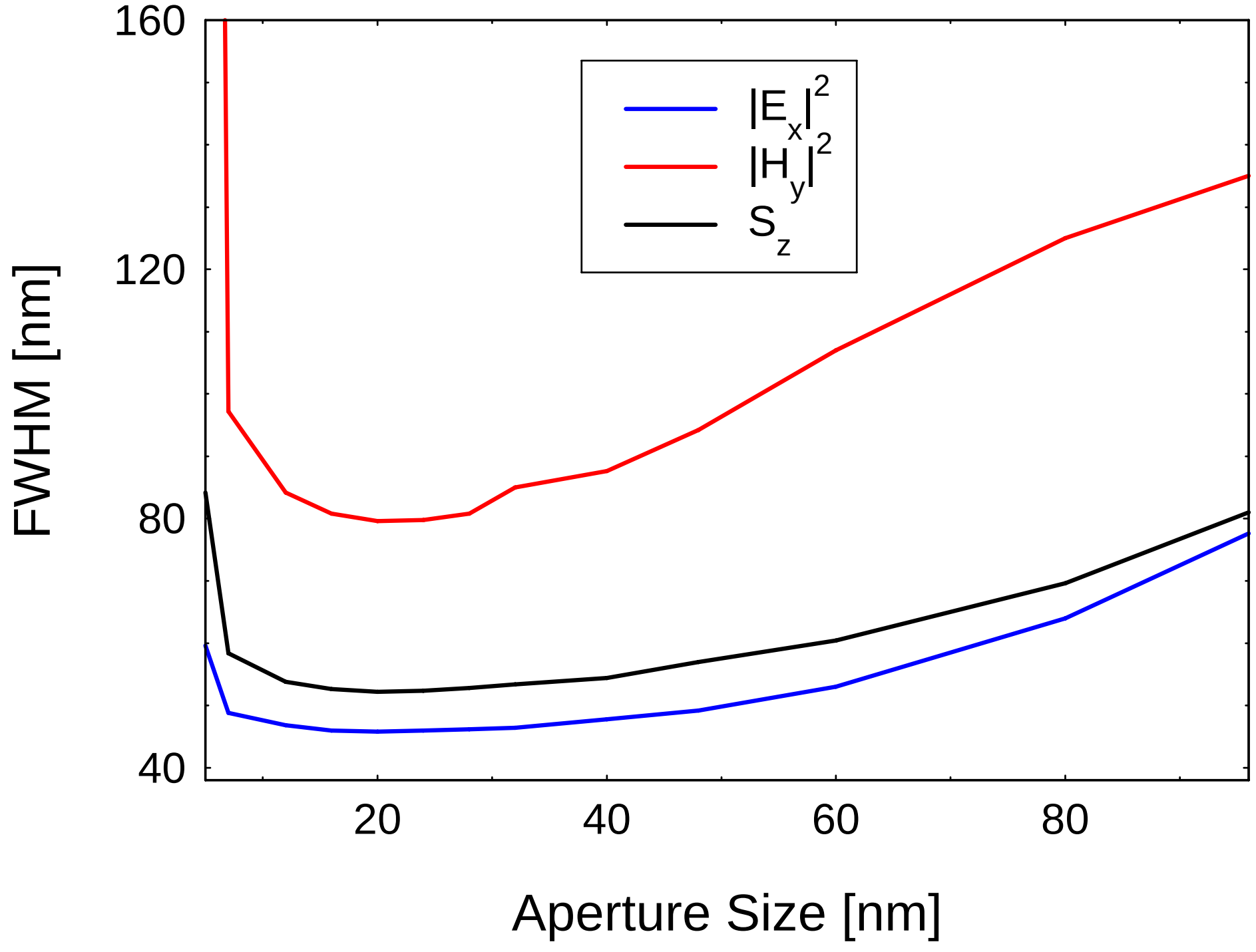

Fig.12b

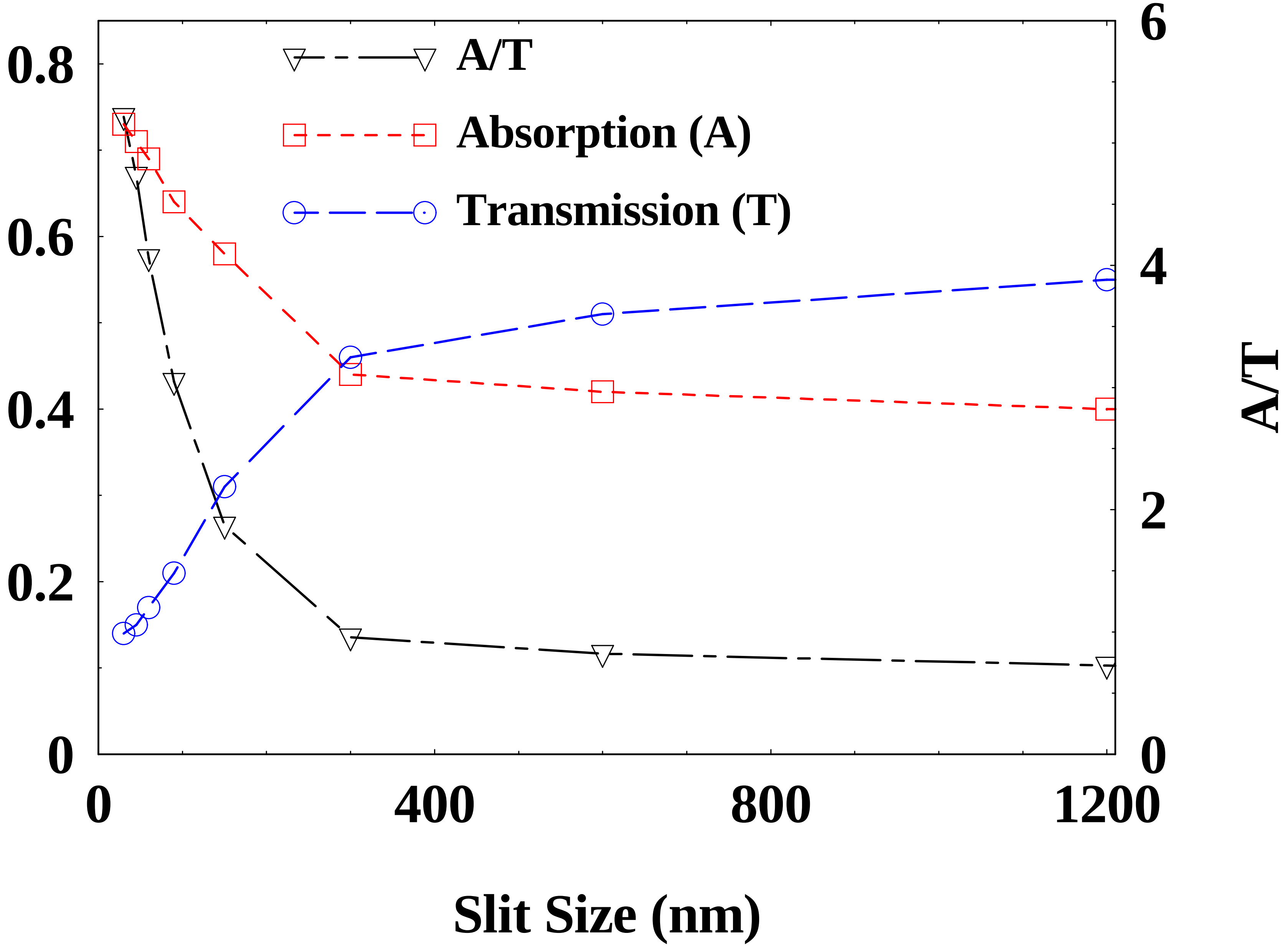

Fig.13

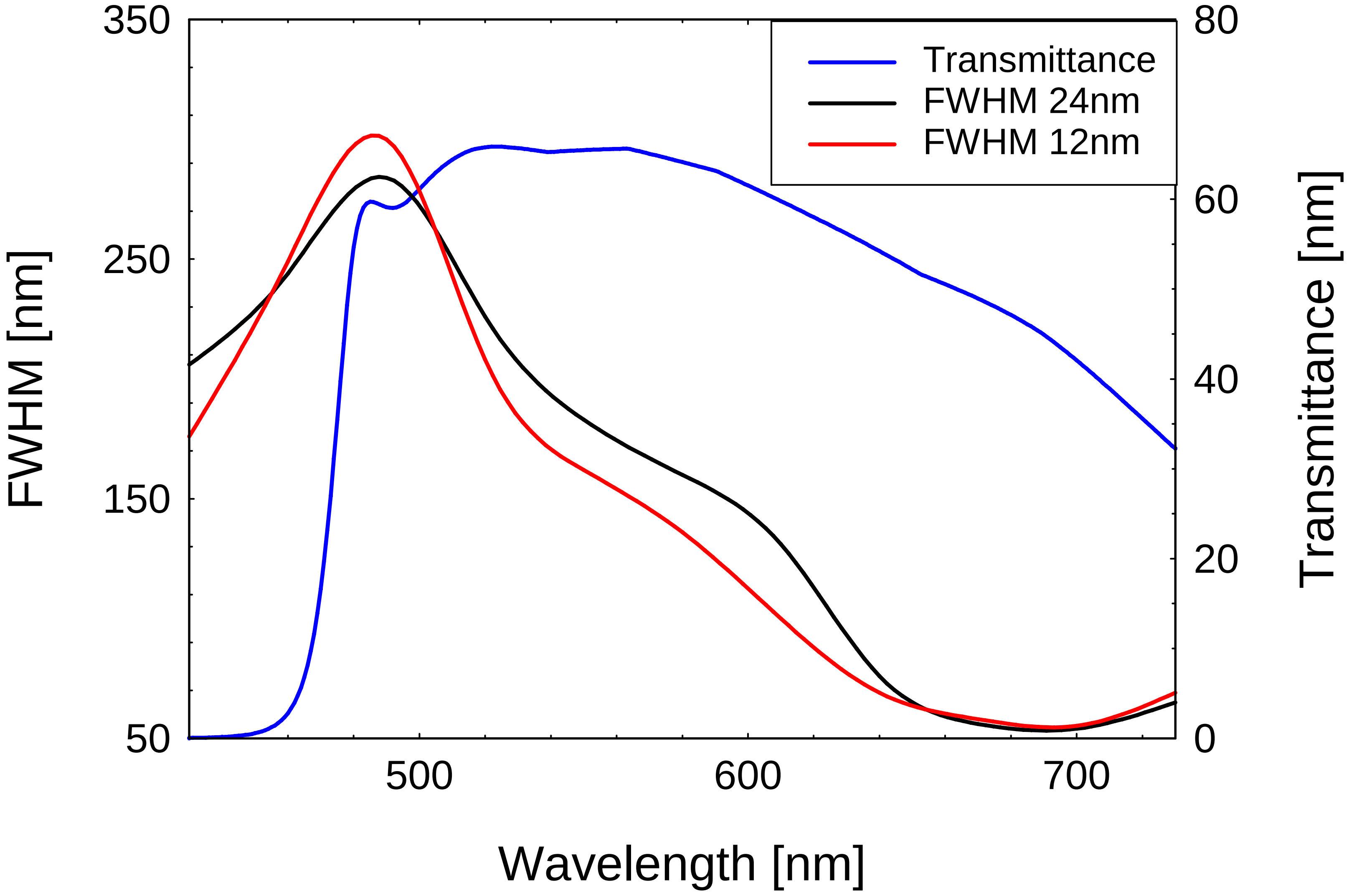

Fig.14

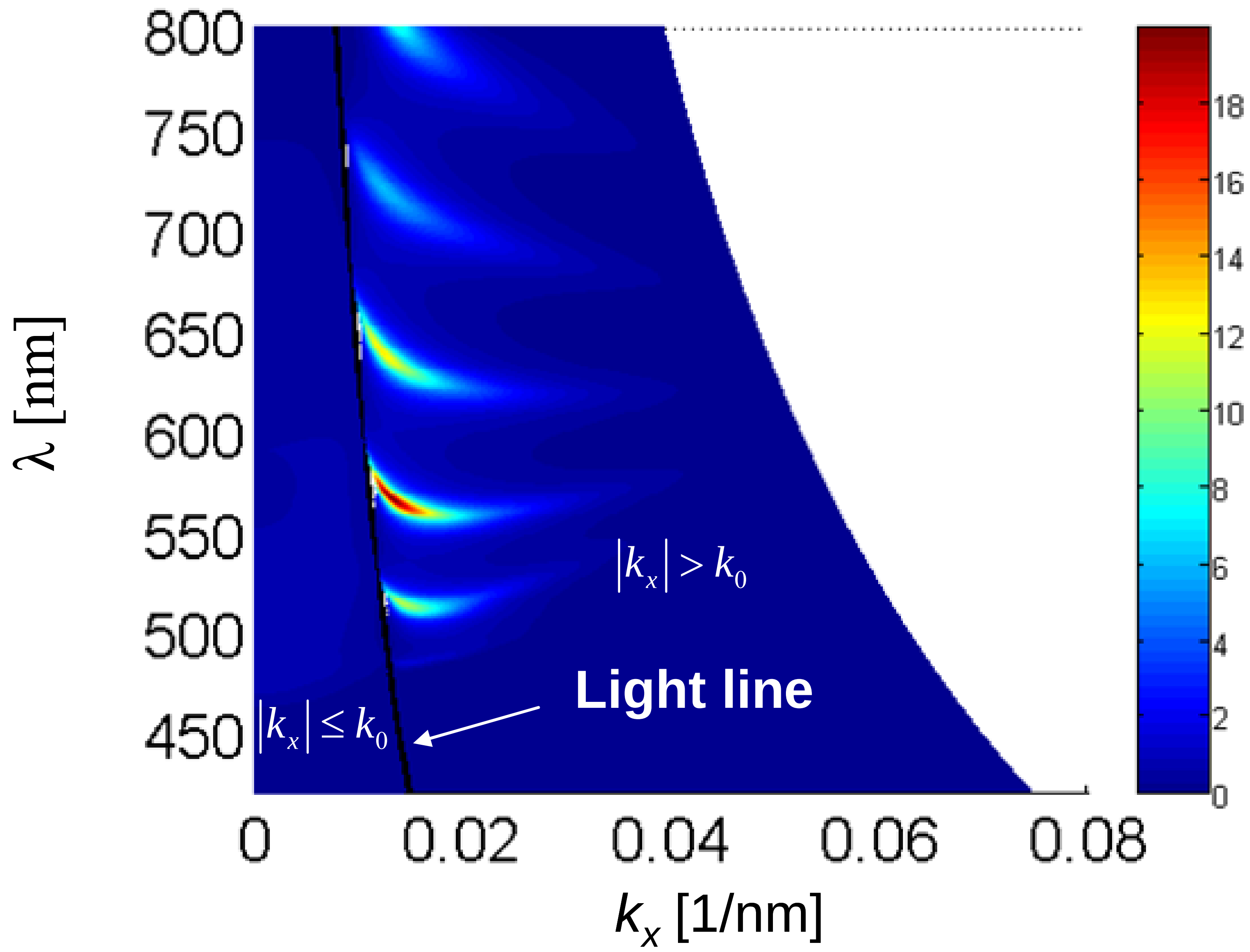

Fig.15

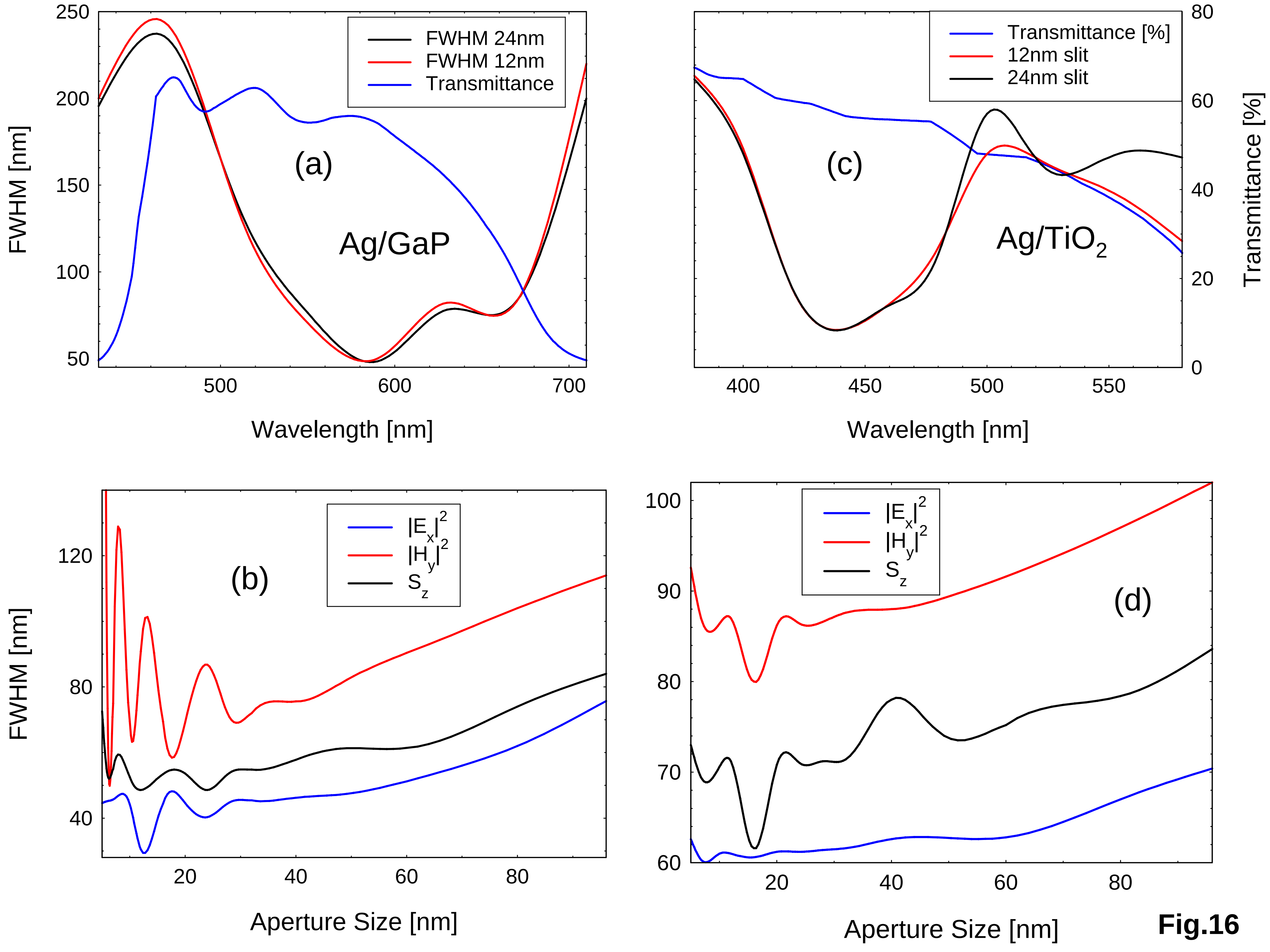

**Fig.16**

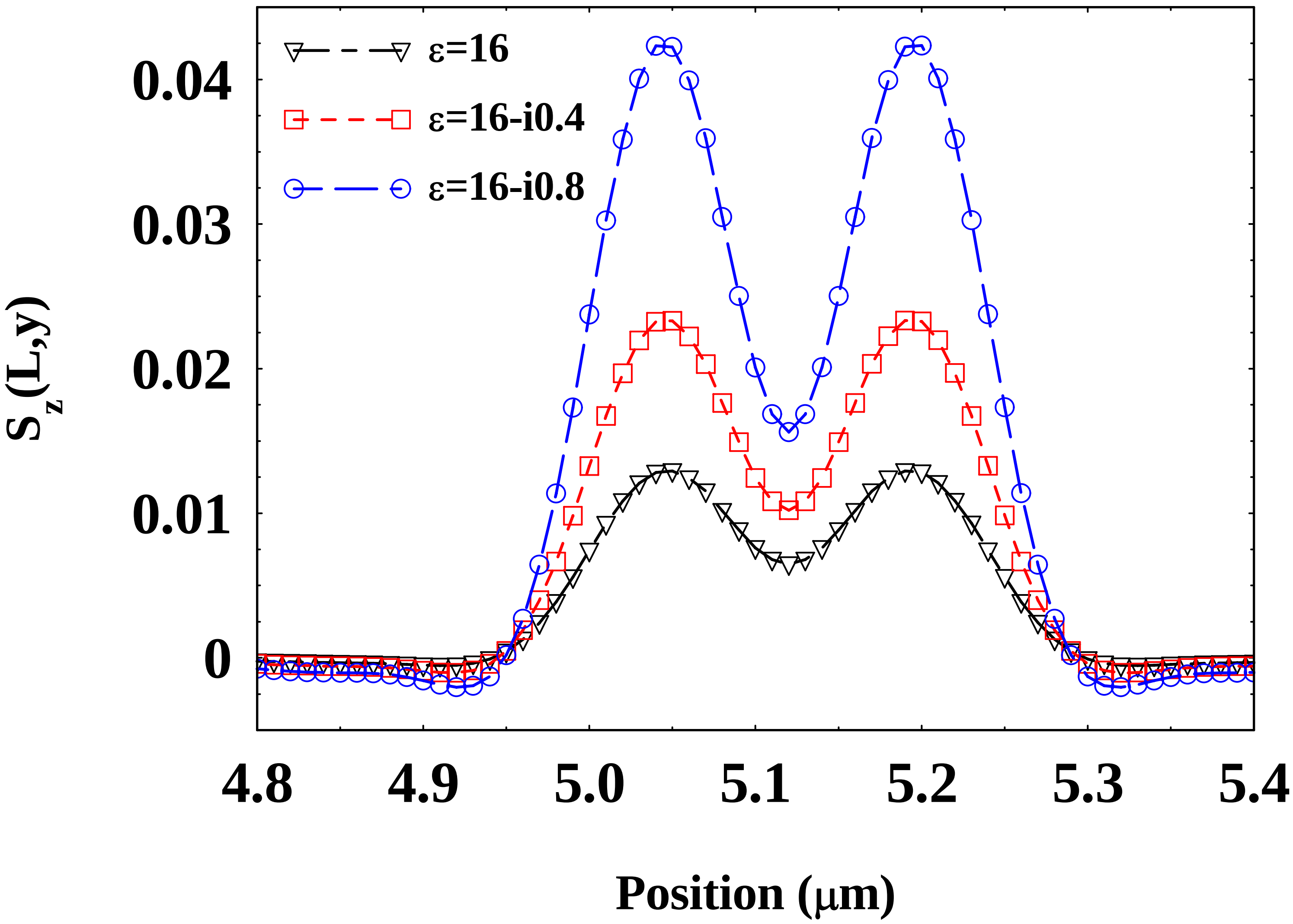

Fig.17